\documentstyle[12pt,equation]{article}
\begin{document}
\pagestyle{empty}
\def\to{\rightarrow}
\def\rs{\mbox{$\sqrt{s}$}}
\def\pt{\mbox{$p_T$}}
\def\ccbar{\mbox{$c \bar c$}}
\def\f2gam{\mbox{$F_2^\gamma$}}
\def\gstarga{\mbox{$\gamma^* \gamma$}}
\def\gamgam{\mbox{$\gamma \gamma$}}
\def\eplem{\mbox{$e^+e^-$}}
\def\glgam{\mbox{$g^\gamma (x,Q^2)$}}
\def\qgam{\mbox{$q^\gamma (x,Q^2)$}}
\def\cgam{\mbox{$c^\gamma (x,Q^2)$}}
\def\fgame{\mbox{$f_{\gamma |e} $}}
\def\fcgam{\mbox{$f_{c|\gamma}$}}
\def\fp1p{\mbox{$f_{P_1|p}$}}
\def\fgp{\mbox{$f_{g/p}$}}
\def\qsq{\mbox{$Q^2$}}
\def\be{\begin{equation}}
\def\ee{\end{equation}}
\def\bea{\begin{eqnarray}}
\def\eea{\end{eqnarray}}
\hfill{LNF--95/020 (P)}
\vskip 0.1 cm
\hfill{MAD--PH--889}
\vskip 0.1 cm
\hfill{May 1995}
\vskip 0.5 cm
\begin{center}
{\large \bf {Inclusive Charm Production at HERA and the Charm Content of the
Photon}\footnote{Talk presented by R.M. Godbole at Photon'95, Sheffield, UK,
April 8-13,1995.}}
\vglue 0.5 cm
Manuel Drees\footnote{Heisenberg Fellow}\\
{\sl
University of Wisconsin, Dept. of Physics, 1150 University Avenue,
Madison, WI 53706, U.S.A.}
\vglue 0.15 cm
and \\
\vglue 0.15 cm
Rohini M. Godbole\footnote{Permanent address:
Physics Department, University of Bombay, Vidyanagari, Bombay - 400 098,
India }\\
{\sl INFN--Laboratori Nazionali di Frascati, P.O.Box 13, I-00044,
Frascati (Roma), Italy}
\end{center}
\vglue 1cm
\begin{abstract}
\noindent
We calculate the contribution to inclusive high tranverse momentum ($p_T$)
charm production at HERA from the excitation of charm in the photon. At large
values of \pt\ the results of such a calculation, in the structure function
language, will be more reliable as it sums the large logs, $\log(p_T^2/m_c^2)$,
as opposed to calculating the contribution of the $2 \to 3 $ subprocess in
fixed order of perturbation theory. We find that this contribution is large
and comparable to the contribution from $\gamma g$ fusion production of charm.
Suitable cuts on the rapidity of the `away-side' large \pt\ jet allow a very
neat separation between the contributions from the excitation process and from
pair-production. We further find that including this excitation contribution
we can reproduce the measured inclusive $D^*$ and $\mu$ cross--sections
measured by the ZEUS and H1 collaborations respectively, in a LO calculation.
\end{abstract}

\newpage
\pagestyle{plain}
\setcounter{page}{1}
Measurements of \f2gam\ in \gstarga\ scattering at the \eplem colliders PEP,
PETRA, TRISTAN and LEP \cite{1} have by now yielded a lot of information on
the parton content of the photon over a wide range of $x$ and \qsq. However,
these measurements give direct information only about the quark content of the
photon. The gluon density \glgam\ is poorly determined as it affects \f2gam\
only through the QCD evolution equations. At the current values of \qsq\ the
charm quark contribution to \f2gam\ is approximated by the quark-parton-model
(QPM) matrix elements for the process $\gamma \gamma^* \to \ccbar$ and
$\gamma^* g^\gamma \to \ccbar$. Through the latter process, the effective
charm content of the photon becomes sensitive to \glgam. At larger values of
\qsq, \cgam\ computed using the massive Altarelli-Parisi (AP) evoultion
equations, is considerably different from the pure QPM predictions \cite{2}. A
study of the charm content of the photon might also help shed some light on
the correct treatment of a heavy parton inside a target. The various different
available parametrisations of \qgam\ and \glgam\ \cite{3} treat the charm
density \cgam\ with varying amount of rigour and care.  It is therefore
interesting to take a phenomenological approach and think of measurements
which will probe \cgam\ directly and hence perhaps also yield information
about \glgam.

One possibility is to study production of single charm in $ep$ collisions via
the excitation processes (the subprocesses being $c^\gamma + q^p \to c + q$
and $c^\gamma + g \to c +g $, here we neglect the contribution coming from
charm in the proton) shown
in fig. \ref{fig1}. This will give rise to a single high$-p_T$ charm particle
whose transverse momentum is balanced by a light $q/g$ jet. Of course the use
of structure functions to compute this process is meaningful only for large
values of the \pt\ of the charm quark. Admittedly for lower values of \pt\ the
more reliable computation will be that of the $2 \to 3$ subprocesses (some of
which are shown in
fig. 2), but at larger values of \pt\ the structure function language sums up
the large $\log(p_T^2/m_c^2)$ terms and hence is more accurate. Another
contribution to the inclusive charm signal comes from \ccbar\ pair production,
via the `direct' $\gamma g$ fusion subprocess as well as the `resolved'
processes, where the balancing high-\pt\ jet is the $c\ (\bar c)$ quark jet.

The excitation contribution of diagrams in fig. \ref{fig1} is given, in the
Weizs\"acker-Williams (WW) approximation, by
\begin{eqalignno}
{d \sigma^{exc} \over dp_T}= 
{\sum_{P_1}}&\int_{z_{min}}^{z_{max}}
\fgame (z)\; dz\; \int_{x_\gamma^{min}}^1 \fcgam (x_\gamma) dx_\gamma
\nonumber\\
& \cdot \int_{x_p^{min}}^1 \fp1p (x_p) dx_p\; {d \hat \sigma \over dp_T}(P_1 +
c^\gamma \to c + P_1),
\label{exc}
\end{eqalignno}
where \fgame, \fcgam\ and \fp1p represent the flux factors of the $\gamma$ in
the electron, charm in the photon and parton $P_1$ in proton respectively;
\begin{eqalignno}
z_{min} &= {\rm max} \{z_{min}^{kin}, z_{min}^{exp}\} \nonumber \\
z_{max} &= {\rm min} \{z_{max}^{exp},1\},
\nonumber
\end{eqalignno}
where $z_{min(max)}^{exp}$ correspond to the experimental cuts on the outgoing
electron (or equivalently the $\gamma$ energy) and $z_{min}^{kin},
x_\gamma^{min}$ and $x_p^{min} $ correspond to the kinematic limits on the
different momentum fractions. The direct contribution to \ccbar\ pair
production is similarly given by
\be
{d \sigma^{pair} \over dp_T} = \int_{z_{min}}^{z_{max}} \fgame (z)
\; dz\; \int_{x_p^{min}}^1 \fgp (x_p) dx_p \;
{d \hat \sigma \over d\pt}(\gamma + g \to c + \bar c),
\label{direct}
\ee
where \fgp\ represents the gluon flux in the proton, and the `resolved'
contribution is given by an expression similar to Eq.(\ref{exc}) where
contributions from all the various subprocesses involving all the different
partons in the photon are to be included. We use LO expressions for all the $2
\to 2$ subprocess cross--sections \cite{3}.

The virtuality $-P^2$ of the exchanged photon in figs. \ref{fig1} and 2
can, in principle, affect its parton content \cite{4}. For the results
presented here we impose the requirement $P^2 < 0.01\ {\rm GeV}^2$ and $0.25
< z < 0.70$, following the cuts used in the experimental study of the
photoproduction of jets \cite{5}. This implies that our expression for the
photon flux factor is given by
\be
\fgame = {\alpha \over {2 \pi z}} [ 1 + (1-z)^2 ] \; \; \ln
\left({0.01\ {\rm GeV}^2 \over {P^2_{min}}} \right) \;\;
- \; {\alpha \over \pi} {{1-z} \over z},
\ee
where
$$
P^2_{min} = m_e^2 {z^2 \over {(1-z)}}.
$$
As a result of the cut on $P^2$ we can neglect the effect of the virtuality
of the $\gamma$ on its parton content.

The results of our computations are shown in fig. \ref{fig3} for various
photon structure function parametrisations DG \cite{6}, LAC\cite{7},
WHIT\cite{2} and different proton structure functions \cite{8}; we use
$\Lambda_{QCD} = 0.4$ GeV for the DG and WHIT parametrizations, and 0.2 GeV
for LAC. We see from this figure that the excitation cross--sections are
indeed comparable to the \ccbar\ production cross--sections. This implies that
while the `resolved' contribution to \ccbar\ pair production is small (as
shown by the long dashed line) for these large values of \pt, the inclusive
charm signal still has a considerable `resolved' component due to the
excitation contribution. Though we do not show them separately here, the
contribution to $\sigma^{exc}$ coming from a gluon in the initial state
dominates over most of the \pt\ range. Apart from the DG parametrisation for
which excitation contributions are about a factor 2 higher than the rest, the
charm excitation cross--section at HERA seems to be fairly independent of both
the photon and the proton structure function parametrisations. Since the DG
parametrisation has $c^\gamma = u^\gamma$ it definitely overestimates the
charm excitation and hence this part of the result is easily understood. In
principle, the other parametrisations of \cgam\ do also look quite different,
both in the small and large $x_\gamma$ region, but the effective $c-$quark
content of the electron, which involves the convolution of this with the WW
function is very similar in the end; this is reflected in the similarity of
the predictions using the LAC and WHIT parametrizations for \cgam. This can be
looked upon as a positive point in that the size of the excitation
contribution to the inclusive charm signal, at large \pt, can be estimated
quite reliably. However, it is also clear that one needs to devise kinematic
cuts to separate the excitation contribution to the inclusive charm signal
from that due to \ccbar\ pair production. (This is also necessary if direct
\ccbar\ pair production is to be used to study the gluon density in the
proton.)

To this end, we next study the kinematic distributions of the decay muon which
is used to tag the charm in the final state and also that of the balancing
(`away--side')
jet. In this calculation we include fragmentation of the $c-$quark into a
charmed meson \`a la Peterson fragmentation function \cite{9} with the
parameter $\epsilon$ as given in ref.\cite{9a}, and use the value 0.1 for the
semileptonic branching ratio of the charmed hadrons. Since, for our \pt\ cut,
\ccbar\ production is dominated by the `direct' process, the real kinematical
difference between the excitation and \ccbar\ contributions to the charm
signal comes from the fact that in the excitation process only a fraction of
the $\gamma$ energy is available for the subprocess, whereas in the `direct'
process all of it goes into the subprocess. As a result, the
direct process on the whole receives contributions from smaller $x_p$ values
than the excitation process does. Hence the $\bar c$ jet in the \ccbar\ case
will have much more negative rapidity than the `light' ($q/g$) jet in the
excitation process (the proton direction is taken as positive $z$ axis); this
is very similar to the corresponding situation with the photoproduction of
jets \cite{3}. On the other hand, the rapidity distributions of the charm
quarks produced in both the excitation and pair production processes are very
similar and hence those of the decay muons also. The kinematic distribution in
$p_T^\mu$ and $y_\mu$ therefore are very similar for both contributions.
Fig. 4 shows the rapidity distribution of the jet balancing the large
\pt\ charm, with the WHIT5 parametrisations of \qgam\ and \glgam\ and MRSD-
for the proton structure function. As we can see very clearly from the figure,
a cut on $y_{jet} < 0.5$ can neatly separate the excitation and the \ccbar\
contribution from each other. The rates presented in the figure include the
semileptonic branching ratio of the charm meson. It should also be mentioned
that these distributions do have some sensitivity to $c^\gamma$, but only for
negative values of $y_{jet}$ where the signal is dominated by the \ccbar\
contribution. The figure also tells us that the signal is healthy even after
these cuts and hence is measurable. For a clear signal one will have to make
an additional cut on $p_T^\mu$ as well but that will affect both the
excitation and the pair production contribution similarly.

Recently both H1 and ZEUS have reported measurements of inclusive charm
production at HERA \cite{10}. ZEUS reports $D^{*}$ production with $p_T(D^*) >
1.7 $ GeV and $|\eta(D^*)| < 1.5$ where $\eta$ is the pseudorapidity with $P^2
< 4\ {\rm GeV}^2$ and $0.15 < z < 0.86$, whereas H1 reports observation of
events with a hard muon with $p_T^\mu > 1.5$ GeV and $30^{\circ} < \theta(\mu)
< 130^{\circ}$. The ZEUS analysis then uses this measured cross--section to
estimate the `total' \ccbar\ cross--section by extrapolating it outside the
measured region and then compare the value so obtained with the QCD NLO
calculations. We attempted instead to reproduce the cross--sections measured
by ZEUS and H1 by using our LO QCD calculations. Since the $p_T$ cut and $m_c$
are comparable it is not clear whether factorisation of the production and
fragmentation of the charm quark is such a good approximation. On the other
hand if we include the fragmentation of the $c-$quark in the final state then
we must include the excitation contribution which corresponds to the
fragmentation of the initial state photon into charm. We therefore run our
Monte Carlo with two different options: In one case (A) we fragment the charm
using the Peterson fragmentation function and include the excitation
contributions whereas in the other case (B) we do not include the
fragmentation of the final state $c$ quark and drop the excitation
contributions as well. Eventually detailed comparisons with measured
transverse momentum and rapidity distributions should reveal which discription
is more appropriate. When comparing with ZEUS results we include a factor of
0.26 which is the probability of a charm quark to fragment into a charged
$D^*$ meson.

\begin{table}[hbt]
\caption{ The $D^*$ and $\mu$ cross--sections measured at HERA by ZEUS and
H1, compared with the LO predictions discussed in the text.}
\begin{center}
\begin{tabular}{|c|c|c|c|c|c|}
\hline
\multicolumn{3}{|c|}{ZEUS ($D^*$)} &\multicolumn{3} {|c|} {H1 ($\mu$) } \\
\cline{1-3} \cline{3-6}
\multicolumn{3}{|c|}{ Data: $32 \pm 7 ^{+4}_{-7} $ nb}
&\multicolumn{3}{|c|}{  Data: $2.03 \pm 0.43 \pm 0.7$ nb} \\
\cline{1-3} \cline{3-6}
\multicolumn{3}{|c|}{A: Frag and $c^\gamma$ included}
&\multicolumn{3}{|c|}{A: Frag and $c^\gamma$ included} \\
\cline{1-3} \cline{3-6}
p str.fn.&$\gamma$ str. fn.&$\sigma$ (nb)& p str.fn.
&$\gamma$ str. fn.&$\sigma$ (nb)\\
\hline
MRSD-&LAC1&27.4&MRSD-&LAC1&1.5 \\
MRSD-&WHIT5&26.0&MRSD-&WHIT5&1.5 \\
MRSD0&WHIT5&21.9&MRSD0&WHIT5&1.48 \\
MRSD-&{\bf DG}&64&MRSD-&{\bf DG}&2.9 \\
\hline
\multicolumn{3}{|c|}{B: No frag and $c^\gamma$ not included}
&\multicolumn{3}{|c|}{B: No frag and $c^\gamma$ not included} \\
\cline{1-3} \cline{3-6}
p str.fn.&$\gamma$ str. fn.&$\sigma$ (nb)& p str.fn.
&$\gamma$ str. fn.&$\sigma$ (nb)\\
\hline
MRSD-&LAC1&26.8&MRSD-&LAC1&1.8 \\
{\bf MRSD0}&WHIT5&15.4&MRSD0&WHIT5&1.5 \\
MRSD-&WHIT5&37.0&MRSD-&WHIT5&2.4\\
MRSD-& DG&33.4&MRSD-&DG&2.3 \\
\hline
\end{tabular}
\end{center}
\label{table1}
\end{table}

Table \ref{table1} gives a summary of our calculations along with the results
reported by the two experimental groups. We find that for the ZEUS data, in
case A the results become less sensitive to the low-$x$ behaviour of the
proton structure function and the excitation contribution actually dominates.
In this case both the MRSD- and MRSD0 for the partons in the proton and WHIT5
or LAC1 partons in the photon reproduce the data whereas DG predicts too big a
cross--section. On the other hand, in case B, the results are more sensitive
to the low-$x$ behaviour of the parton densities and we find that we can
reproduce the cross--section only for a steeply rising gluon density. In this
case all the photonic parton densities combined with MRSD- for the proton are
acceptable whereas MRSD0 gives answers smaller by a factor 2 for all
reasonable choices of the momentum scale as well as the photonic parton
densities. For the H1 sample, the $p_T^\mu$ cut means that the produced charm
quark has much higher $p_T$ than for ZEUS. Again, inclusion of fragmentation
reduces the sensitivity to the low-$x$ behaviour of the gluon in the proton
and results for various combinations of partons in the photon and proton are
almost the same. As before, the DG parametrisation predicts a large
cross--section, 2.9 nb, but it is not inconsistent with the data.

In conclusion we have studied the contribution to the inclusive charm signal
from the excitation of charm in the photon. We find the rates to be comparable
to the contribution coming from \ccbar\ pair production. Due to the
convolution with \fgame\ the sensitivity of $\sigma^{exc}$ of Eq.(\ref{exc}) to
the region of large $x_\gamma$, where the various parametrisations for \cgam\
differ most, is reduced. This means that this `resolved' background to the
large \pt, inclusive charm signal coming from the `direct' process can be
predicted quite reliably. Making a cut on the `away-side' jet-rapidity allows a
separation of these two contributions. Note that charm excitation events
should contain a second `spectator' charm (anti--)quark in the photon remnant,
which has a large, negative rapidity (close to the electron beam direction).
In contrast, \ccbar\ pair production events should contain a second high$-p_T$
charm (anti--)quark. In both cases the presence of a second charmed particle
should be visible in some fraction of the events; this could be used as a
cross--check of the relative sizes of the two contributions to the inclusive
charm signal. Even after all the cuts and folding of the cross--sections with
various branching fractions, the rates for inclusive charm production are
large and easily measurable. Further, a LO computation including the
excitation contributions gives results comparable to the recent measurements
of the inclusive charm signal by ZEUS and H1. On the theoretical side, it
would be interesting to compare our results with explicit $2 \to 3$ subprocess
calulations and see at what values of \pt\ do they match.

\vspace*{0.5cm}
\noindent
{\bf Acknowledgements:}
The work of M.D. was supported in part by the U.S. Department of Energy under
grant No. DE-FG02-95ER40896, by the Wisconsin Research Committee with funds
granted by the Wisconsin Alumni Research Foundation, as well as by a grant
from the Deutsche Forschungsgemeinschaft under the Heisenberg program. R.M.G.
wishes to acknowledge research grant no. 03(745)/94/EMR--II from the Council
for Scientific and Industrial Research. We would like ot thank the Department
of Science and Technology (India) and the organisers of WHEPP(III) held in
India, where this project was initiated.


\begin{thebibliography}{99}
\bibitem{1}
For the recent measurements see:
D. Morgan, R. Pennington and M.R. Whalley, J. Phys. {\bf G20}, A1 (1994);
OPAL collaboration, R. Akers et al, Z. Phys. {\bf C 61} (1994) 199;
B. Kennedy, these proceedings;
TOPAZ collaboration, Phys. Lett. {\bf B 332} (1994) 477 ;
AMY collaboration, T. Nozaki, these proceedings.
\bibitem{2} K. Hagiwara, M. Tanaka, I. Watanabe and T. Izubuchi, Phys. Rev.
{\bf D 51} (1995) 3197.
\bibitem{3} For a summary and further references see,
M. Drees and R.M. Godbole, Pramana {\bf 41} (1993) 83.
\bibitem{4} See, for example, M. Drees and R.M. Godbole, Phys. Rev. {\bf D 50}
(1994) 3124.
\bibitem{5} H1 collaboration, I. Abt et al, Phys. Lett. {\bf B 314} (1993) 436.
\bibitem{6} M. Drees and K. Grassie, Z. Phys. {\bf C28} (1985) 451.
\bibitem{7} H. Abramowicz, K. Charchula and A. Levy, Phys. Lett. {\bf B 269}
(1991) 458.
\bibitem{8} A. D. Martin, R.G. Roberts and W.J. Stirling, Phys. Lett.
{\bf B 306} (1993) 145; Erratum ibid. {\bf B 309} (1993) 492.
\bibitem{9} Peterson et al, Phys Rev. {\bf D 27} (1983) 105.
\bibitem{9a} Review of Particle Properties, Phys. Rev.{\bf D 50} (1994) 133.
\bibitem{10} ZEUS collaboration, M. Derrick et al, DESY 95-013;
U. Karshon for the ZEUS collaboration, these proceedings;\\
C. Kleinwort for the H1 collaboration, talk given at {\it ICHEP} Glasgow,
August 1994, DESY report 94--187.
\end{thebibliography}
\end{document}